\documentclass[dvips, twocolumn, showpacs, preprintnumbers, amsmath, amssymb]{revtex4}
\usepackage[english]{babel}
\usepackage{graphics} 
\usepackage{graphicx}% Include figure files
\usepackage{amsmath, amsfonts, amssymb}
\usepackage[latin1]{inputenc}

\begin{document}

%---------------------------------------------------------------------------------------
\title{Cluster evolution in steady-state two-phase flow in porous media}
%---------------------------------------------------------------------------------------
\author{Thomas Ramstad}
\email[Thomas.Ramstad@phys.ntnu.no]{}
\affiliation{Department of Physics, Norwegian University of Science and
Technology, N--7491 Trondheim, Norway}

\author{Alex Hansen}
\email[Alex.Hansen@ntnu.no]{}
\affiliation{Department of Physics, Norwegian University of Science and
Technology, N--7491 Trondheim, Norway}
%---------------------------------------------------------------------------------------
\date{\today}
%---------------------------------------------------------------------------------------
\begin{abstract}
We report numerical studies of the cluster development of two-phase flow in a steady-state environment of
porous media.
This is done by including biperiodic boundary conditions in a two-dimensional flow simulator. 
Initial transients of wetting and non-wetting phases that evolve before steady-state has occurred, 
undergo a cross-over where every initial patterns are broken up. For flow dominated by capillary effects with
capillary numbers in order of $10^{-5}$, we find that around a critical saturation of
non-wetting fluid the non-wetting clusters of size $s$ have a power-law distribution $n_s \sim s^{-\tau}$ with the exponent 
$\tau = 1.92 \pm 0.04$ for large clusters. This is a lower value than the result for ordinary percolation.
We also present scaling relation and time evolution of the structure and global pressure.
\end{abstract}
%---------------------------------------------------------------------------------------
%---------------------------------------------------------------------------------------
\maketitle
%---------------------------------------------------------------------------------------

\section{Introduction}\label{1} 

The complex nature of multi-phase flow in porous media has acted as 
motivation for extensive studies in the recent years. Different types
of fluid displacements in porous media play important roles in natural 
processes, many of them of large practical importance such as 
oil recovery, soil mechanics and hydrology. Large quantities of petroleum and water
resources are hidden in fractured and porous stones and that is of
vast economic and social importance.
 
Two-phase displacement in porous media has been studied through seminal experimental
work \cite{cw85, mfj85, ltz88}, numerical simulations \cite{kl85, bk90, r90} and theoretical
work including statistical models and differential equations \cite{ww83, bc98}. Dynamics of 
instabilities in immiscible two-phase flow 
controlled by the interplay between viscous and capillary forces, give 
rise to complex pattern formations. 

The traditional way of describing those situations has been through either drainage or
imbibition. That will give rise to transient effects of front propagation 
such as invasion percolation (IP),
viscous fingering and stable front displacement. The common feature of these
effects are that they are out of equilibrium, and the IP regime will eventually even end with a static regime
with immobile structures due to large capillary barriers. 
Most experiments and simulations in this area deal with systems where one and only one fluid is injected into another
until the injected fluid reaches the sink of the system. 
In natural reservoirs of fluid in porous media the situation is, however, dynamic and governed 
by the interplay and competition 
between drainage and imbibition. This is best described as a steady-state regime inside a representative
volume element where both drainage and imbibition occurs. Pure drainage or pure imbibition are pore-scale concepts, and can therefore 
not alone be scaled up in a steady-state situation. 

In this paper we examine distribution of cluster formations in a pore scale network under
both steady-state and transients. This study contains both quantitative and
qualitative results of cluster distribution and dynamic evolution of the phases. 
The model used in our simulations is mainly similar to that proposed by Aker {\it et al} \cite{amhb98} 
with later generalizations by Knudsen {\it et al} \cite{ref2} to include {\it biperiodic} boundary conditions. 
The boundary conditions in the flow direction lets the fluid that leaves the system enter the system again at the inlet of the
model in a seamless manner. 

This ables us to investigate steady-state flow which inflects the situation deep inside reservoirs.  With 
both an invading front of non-wetting fluid and wetting imbibition the system will give rise to a blending of structure formations. 
These formations are formed by the interplay of different effects like fragmentation of large
clusters, merging of smaller clusters and a diffusion of fragments. 

This paper is organized in the following way. In Section~\ref{2} we describe outlines of the model
used for our simulations, and in Section~\ref{3} we sketch the initial conditions used followed by 
a discussion of the qualitative behaviour of initial transients.

Further, we discuss quantitative features of the simulations in Section~\ref{4}, prior to a
scaling analysis.

\section{Model}\label{2}
The main statistical models that reproduce the basic domains of porous flow 
belong to the family of growth
models that obey the Laplacian equation $\nabla^2 P = 0$ where $P$ is a pressure
field and with an interfacial growth rate $q \propto \nabla P$. 

The simulations used in this work is based upon a basis model of
disordered media, that consists of tubes orientated 
45 degrees to the overall flow direction \cite{amhb98, ref2}. The network has a coordination number of $4$, 
and the arrangement of the tubes is square planar. The tubes consist of both the pore volume
and the throat volume. The points where the tubes meet are referred to as nodes. 
Randomness is incorporated in the model by allowing
the length of the tubes $d_{ij}$ to be chosen randomly within $30\%$ of the mean length 
i.e. the lattice constant $d$, and the radii to be chosen from a flat distribution 
$r \in (0.1d , 0.1d + 0.3d_{ij})$.

Distributions of fluid that we report are
highly dependent of saturation of phases and ${\rm Ca}$. 
The capillary number
%%%%%%%%%%%%%%%%%%%%%%%%%%%%%%%%%%%%%%%%%%%%%%%%%%%%%%%%%%%%%%%%%%%%%%%%%%%%
\begin{equation}
{\rm Ca} = \frac{\mu Q_{tot}}{\gamma \Sigma} \ ,
\end{equation} 
%%%%%%%%%%%%%%%%%%%%%%%%%%%%%%%%%%%%%%%%%%%%%%%%%%%%%%%%%%%%%%%%%%%%%%%%%%%%
\noindent denotes the ratio between viscous and capillary forces where $\mu$ is 
the largest viscosity rate of the two fluids, $\gamma$
is the surface tension, $\Sigma$ denotes the total input surface of the system and
$Q_{tot}$ is the global flow rate in the system.
The patterns that form during invasion of a non-wetting fluid into a defending wetting one
are characterized mainly into viscous fingering, capillary fingering or stable displacement depending
on ${\rm Ca}$ and the viscosity ratio of the defending $\mu_1$ and invading phase $\mu_2$ given by 
%%%%%%%%%%%%%%%%%%%%%%%%%%%%%%%%%%%%%%%%%%%%%%%%%%%%%%%%%%%%%%%%%%%%%%%%%%%%
\begin{equation}
M = \frac{\mu_{\rm nw}}{\mu_{\rm w}} \ .
\end{equation} 
%%%%%%%%%%%%%%%%%%%%%%%%%%%%%%%%%%%%%%%%%%%%%%%%%%%%%%%%%%%%%%%%%%%%%%%%%%%%

Our system of tubes is filled with two fluids, wetting and non-wetting respectively, each
having a certain fraction of the total volume denoted non wetting 
saturation $S_{\rm nw}$ and wetting saturation $S_{\rm w}$. The fluids are separated through many
menisci. The model does not include film flow and only one bubble is allowed inside
one tube at the time. This means that if more than two menisci are created within a tube during
the same time step, they are subsequently collapsed into two menisci.

Since the fluids are immiscible, the meniscus gives rise to a capillary pressure
within the tube. The capillary pressure is dependent on both the surface tension $\gamma$ and the
radius of the tube given by the Young-Laplace law

%%%%%%%%%%%%%%%%%%%%%%%%%%%%%%%%%%%%%%%%%%%%%%%%%%%%%%%%%%%%%%%%%%%%%%%%%%%%%%%%%%%%%%%%
\begin{equation}
p_c = \frac{2\gamma}{r} \cos \theta \ ,
\label{pc}
\end{equation}
%%%%%%%%%%%%%%%%%%%%%%%%%%%%%%%%%%%%%%%%%%%%%%%%%%%%%%%%%%%%%%%%%%%%%%%%%%%%%%%%%%%%%%%%

\noindent where $\theta$ is the wetting angle, i.e. the contact angle between the wetting 
phase and the wall of the tube. In our model, the tubes are considered to be cylindrical
with respect to permeability, but hour-glass shaped with respect to capillary
pressure. The relation for the capillary pressure reads

%%%%%%%%%%%%%%%%%%%%%%%%%%%%%%%%%%%%%%%%%%%%%%%%%%%%%%%%%%%%%%%%%%%%%%%%%%%%%%%%%%%%%%%%
\begin{equation}
p_c = \frac{2\gamma}{r} \left( 1 - \cos(2\pi x)\right) \ ,
\label{pc2}
\end{equation}
%%%%%%%%%%%%%%%%%%%%%%%%%%%%%%%%%%%%%%%%%%%%%%%%%%%%%%%%%%%%%%%%%%%%%%%%%%%%%%%%%%%%%%%%

\noindent where $x$ is the normalized position of the meniscus running from $0$ to $1$.
The local flow rate $q$ of a tube follows the Washburn equation for capillary flow 
\cite{dullien}, 

%%%%%%%%%%%%%%%%%%%%%%%%%%%%%%%%%%%%%%%%%%%%%%%%%%%%%%%%%%%%%%%%%%%%%%%%%%%%%%%%%%%%%%%%
\begin{equation}
q =  -\frac{k}{\mu_{\rm{eff}}} \frac{\pi r^2}{d} (\Delta p - p_c) \ ,
\label{q}
\end{equation}
%%%%%%%%%%%%%%%%%%%%%%%%%%%%%%%%%%%%%%%%%%%%%%%%%%%%%%%%%%%%%%%%%%%%%%%%%%%%%%%%%%%%%%%%

\noindent where $p_c$ is given by Eq. \eqref{pc2} and the permeability $k = r^2/8$. This
permeability is taken directly from Hagen-Poiseulle flow in cylindrical tubes. We define the
effective viscosity as

%%%%%%%%%%%%%%%%%%%%%%%%%%%%%%%%%%%%%%%%%%%%%%%%%%%%%%%%%%%%%%%%%%%%%%%%%%%%%%%%%%%%%%%%
\begin{equation}
\mu_{\rm eff} =  \mu_{\rm nw}x_{\rm nw} + \mu_{\rm w}(1-x_{\rm nw}) \ ,
\label{mueff}
\end{equation}
%%%%%%%%%%%%%%%%%%%%%%%%%%%%%%%%%%%%%%%%%%%%%%%%%%%%%%%%%%%%%%%%%%%%%%%%%%%%%%%%%%%%%%%%
\noindent with respect to the position of the non-wetting meniscus $x_{\rm nw}$.
The pressure difference between the inlet and outlet of a tube is $\Delta p$.
 
To solve the transport equations we use the property that no net volume can be stored in the
nodes connecting tubes. The net flux through a node is therefore zero and this gives a set of
linear equations to solve. To find the positions of menisci in the tubes we integrate Eq.
\eqref{q} forward with a predefined time step, $\Delta t$, according to an explicit Euler scheme.
Inside a tube one meniscus moves with a front speed determined from
the local flow rate $q$, but when a meniscus reaches the end of a tube it is further distributed
among the other tubes with ingoing flux. Here it is crucial that volume is conserved.
Further details can be found in ref. \cite{amhb98, ref2}. 

The majority of existing models are quasi static and therefore apply only when the capillary
forces are dominant. This model, however, accounts for dynamic effects and is therefore also
capable of handling fast flow or flow with little surface tension, hence a wider range of the capillary 
number ${\rm Ca}$.

\subsection{Boundary conditions}\label{21}
Most experimental setups and simulations are done on systems that are out
of steady-state. They mainly deal with 
systems where one fluid is injected into another defending fluid and ends when the
invading fluid as reached the sink of the model. 

\begin{figure*}
  \centering
  \includegraphics[width = 0.46\linewidth]{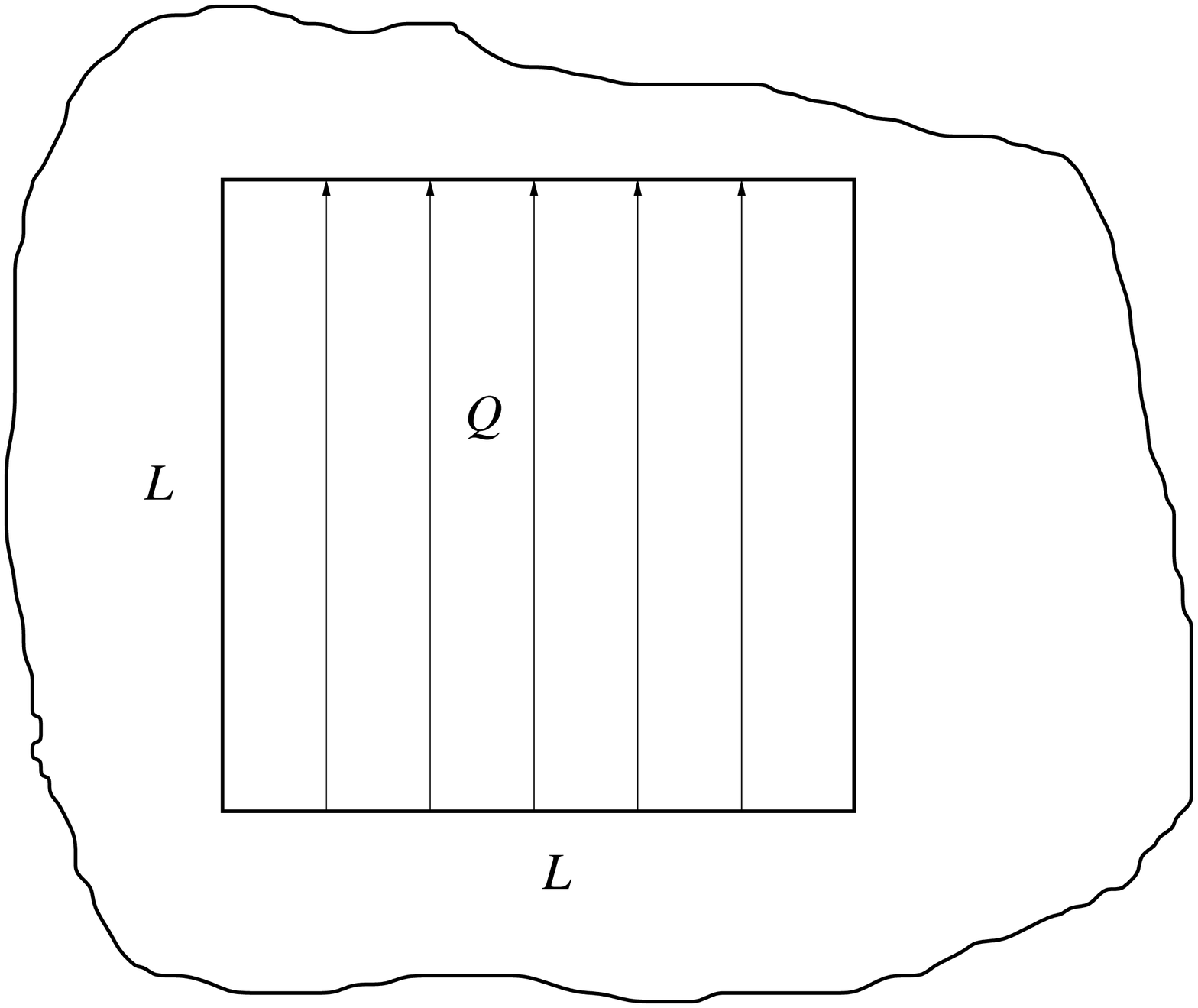}
  \hfill
  \includegraphics[width = 0.46\linewidth]{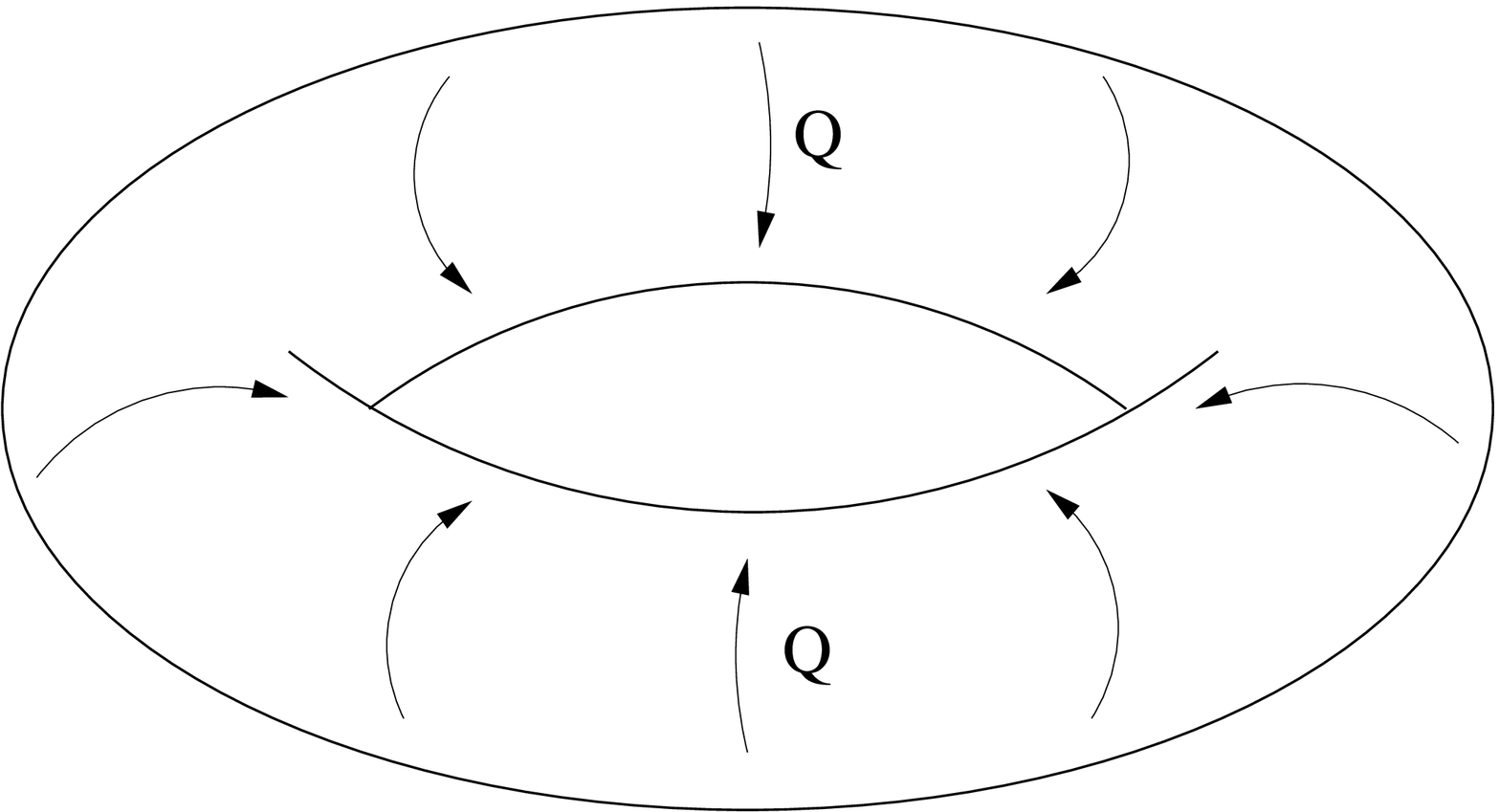}
  \caption{Schematic figure that shows the interpretation the {\it biperiodic boundary conditions} 
    mean to our simulations. In order to 
    model a situation deep inside a natural reservoir in steady state, 
    we must consider the system size as a small part of a greater
    system. Statistically, it is the same fluid-configurations entering the element as are leaving it. 
    This is illustrated in the figure on the left hand side. The way we 
    consider our system is shown on the figure to the right where the two-dimensional system
    is mapped on to a three-dimension torus.}
  \label{setup}
\end{figure*}

In order to simulate steady-state flow, we use {\it biperiodic boundary conditions}. This is done by 
connecting the inlet and outlet row, 
placing the system on a surface with one extra dimension. In two dimensions
this is easily pictured by placing it on a three-dimensional torus as shown in Fig.~\ref{setup}. 

To make the system develop in time, the global pressure field $\Delta P$ is applied over the row where
the original inlet and outlet rows meet. This is done by the use of {\it ghost sites} \cite{r}
where the global pressure is applied.

With biperiodicy in the boundary conditions we will have two invading processes that compete. 
One being the original
invading process as described above with a non-wetting phase penetrating a wetting phase, but 
there will also be a wetting
phase displacing a non-wetting phase since the wetting fluid imbibition from the system will 
enter it at the inlet row.

%%%%%%%%%%%%%%%%%%%%%%%%%%%%%%%%%%%%%%%%%%%%%%%%%%%%%%%%%%%%%%%%%%%%%%%%%%%%%%%%%%%%%%%%%%%%%%%%%
%%%%%%%%%%%%%%%%%%%%%%%%%%%%%%%%%%%%%%%%%%%%%%%%%%%%%%%%%%%%%%%%%%%%%%%%%%%%%%%%%%%%%%%%%%%%%%%%%% 

\section{Initial transients}\label{3}

%%%%%%%%%%%%%%%%%%%%%%%%%%%%%%%%%%%%%%%%%%%%%%%%%%%%%%%%%%%%%%%%%%%%%%%%%%%%%%%%%%%%%%%%%%%%%%% 
\begin{figure*}
  \begin{minipage}{0.46\linewidth}
    \includegraphics[width = 0.9\linewidth]{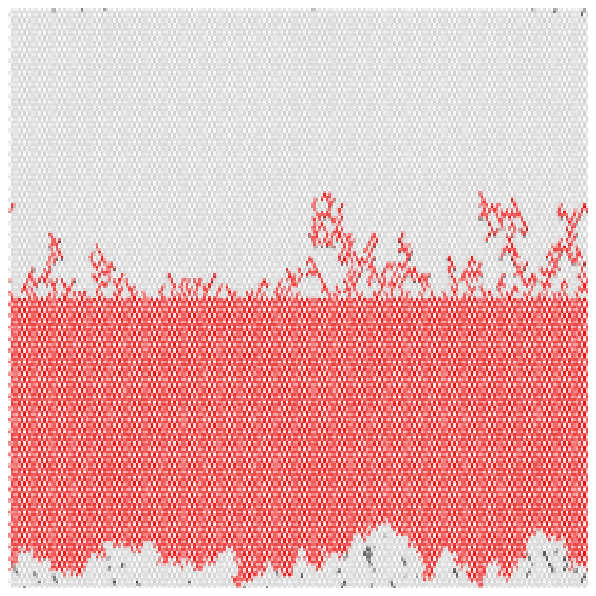}
    \vspace{2mm} 
  \end{minipage}
  \begin{minipage}{0.46\linewidth}
    \includegraphics[width = 0.9\linewidth]{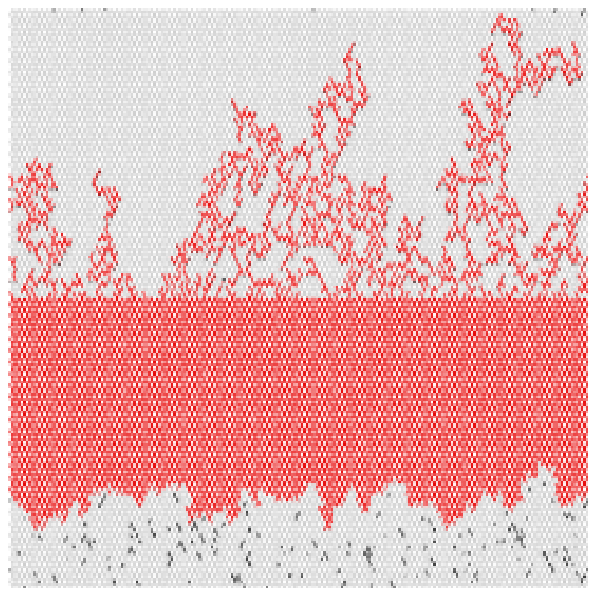}
    \vspace{2mm}
  \end{minipage}
  \begin{minipage}{0.46\linewidth}
    \includegraphics[width = 0.9\linewidth]{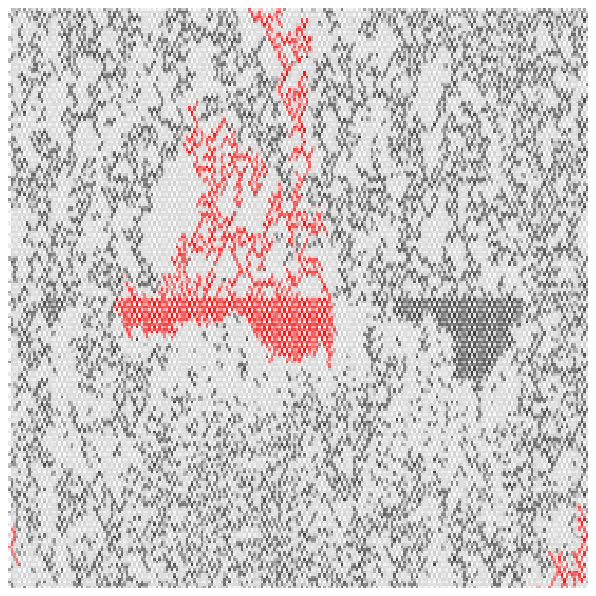}
  \end{minipage}
  \begin{minipage}{0.46\linewidth}
    \includegraphics[width = 0.9\linewidth]{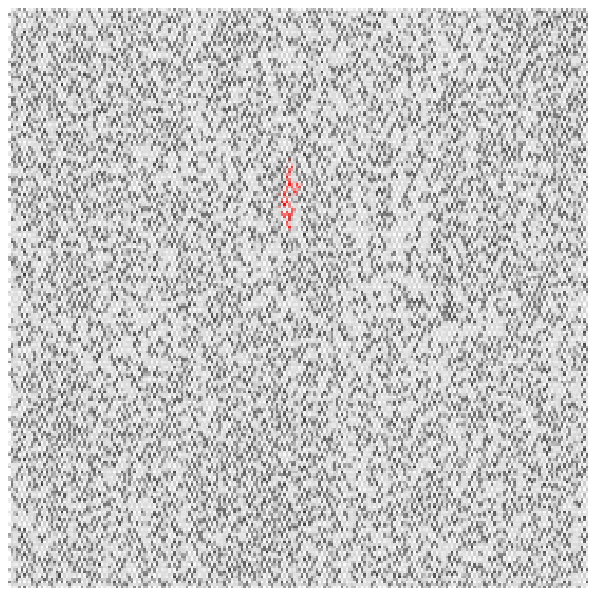}
  \end{minipage}
  \caption{(Color online) Four different stages of a low ${\rm Ca}$ flow configuration with viscosity match $M = 1.0$. ${\rm Ca} = 3.2\times10^{-5}$ 
    implies a capillary-dominant regime. The different snapshots of the flow patterns are taken at $t = 5000\rm{s}$, 
    $t = 10000\rm{s}$, $t = 15000{\rm s}$ and $t = 25000\rm{s}$. The system size is $128\times128$ with a $S_{\rm{nw}} = 0.5$.
    The largest connected cluster is colored red in the online version of the article. The two first snapshot corresponds to the first regime of the pressure in 
    Fig.~\ref{pressure}. In the third snapshot 
    some of the initial configuration is still left and the pressure is building up according to Fig.~\ref{pressure}. 
    The last one is in the steady state regime where every initial structure has vanished.}
  \label{lowCa}
\end{figure*}
%%%%%%%%%%%%%%%%%%%%%%%%%%%%%%%%%%%%%%%%%%%%%%%%%%%%%%%%%%%%%%%%%%%%%%%%%%%%%%%%%%%%%%%%%%%%%%%
In this section we describe the qualitative behaviour of our system as it evolves. As time 
progresses, the dynamic evolution in our model will give rise to different transient effects
that eventually will lead to a steady-state configuration. These transients are dependent on
the control parameters in our system.

Typical parameters that are fixed during our simulations are surface tension 
$\gamma$ and viscosity $\mu$. We keep $\gamma = 30.0 \ {\rm mN} / {\rm m}$ and 
$\mu_{\rm{nw}} = \mu_{\rm{w}} = 0.1 \ {\rm Pa} \ {\rm s}$ which gives the viscosity ratio $M = 1$. 
In order to set the capillary number, we control the overall flow rate $Q_{{\rm tot}}$.

In this paper we consider networks of sizes up to $1024\times 1024$ nodes. We
wish to analyze the distributions of clusters with different ${\rm Ca}$ and different system sizes.

At the beginning of the simulations, the network is segregated into one part of non-wetting fluid and
one part containing only wetting fluid. At the start of the simulation the non-wetting fluid will 
start to invade the wetting fluid part. Depending on the capillary number
it will either start a viscous fingering behaviour for high ${\rm Ca}$ or an IP process when the 
capillary forces dominate. The latter case is visualized in Fig.~\ref{lowCa} where the drainage takes
place in the upper part of the system. In the early stages presented in the figure, a ``belt'' of
non-wetting (black) fluid spans the system in horizontal direction. This ``belt'' is an initial 
configuration chosen by us and will be referred to later in this paper.

On the other hand, wetting fluid will enter the original non-wetting region from below pushing the 
non-wetting fluid in the advancing direction. This displacement of non-wetting fluid in favour of the wetting
happens in a piston like manner. There are some trapped regions left of non-wetting fluid, but these
are smaller than the trapped wetting clusters in the other end of the model. Since the non-wetting front fingers advance
more rapidly than the wetting, more compact propagating front \cite{mrc91}, it will sooner or later catch up with the 
front of the advancing wetting fluid.
At this point there will be a competition between the two different advancing fronts. 

When the system reaches steady-state there will be a mixing of the two phases. This mixing is dependent both of
${\rm Ca}$ and the saturation of fluids. If the difference in
the saturation of fluids is large one of the phases will span the system entirely, in steady-state, with a large
cluster. How large this difference needs to be in order to get a spanning cluster, depends on the capillary
forces and the structure of the network i.e. the porosity. If the capillary forces dominate over the viscous 
ones, the single phase pressure $P_s$ according to Darcy's law

%%%%%%%%%%%%%%%%%%%%%%%%%%%%%%%%%%%%%%%%%%%%%%%%%%%%%%%%%%%%%%%%%%%%%%%%%%%%%%%%%%%%%%%%
\begin{equation}
\frac{Q}{\Sigma} = \frac{k}{\mu} \frac{\Delta P_{\rm{s}}}{L} \ ,
\label{darcy}
\end{equation}
%%%%%%%%%%%%%%%%%%%%%%%%%%%%%%%%%%%%%%%%%%%%%%%%%%%%%%%%%%%%%%%%%%%%%%%%%%%%%%%%%%%%%%%%
\noindent where $L$ is the system length along the single phase gradient $\Delta P_s$, will
be significantly smaller than the total pressure gradient. In this case if there is a spanning
cluster of one fluid it will be much easier for the fluid to move inside this cluster because of
the high capillary pressure threshold in non-occupied tubes. As a consequence once a spanning
cluster evolves, the total pressure will drop and one phase stays immobile or trapped. 

Since the total flux is constant and the capillary pressure will vary with time as
menisci configurations change, the global pressure will fluctuate. However, when the system
reaches steady-state the pressure will fluctuate around a mean value. How long it takes before our
system reaches steady-state is a matter of the initial configuration of the two fluids. 
As the pressure relaxes around a mean value the fractional flow of each fluid will
settle. The non wetting fractional flow is defined as $F_{\rm{nw}} = Q_{\rm{nw}}/Q_{\rm{tot}}$ and the
wetting fractional flow is $F_{\rm{w}} = Q_{\rm{w}}/Q_{\rm{tot}}$. 

It has been shown in previous work \cite{ref2, ref3} that $F_{\rm{nw}}$ depends on both
the capillary number and the saturation $S_{\rm{nw}}$ of the non-wetting fluid. For high values of
${\rm Ca}$, the curve of $F_{\rm{nw}}$ will behave almost linearly with respect to $S_{\rm{nw}}$, while
in the case of very small ${\rm Ca}$ we will get regimes where one fluid is almost immobile. This feature 
was also found to be independent of system size.

For steady-state flow we are only able to obtain a totally immobile regime or a dynamic regime where clusters
rearrange themselves continuously. The 
first situation can be characterized as traditional invasion percolation where capillary forces dominate. 
When a percolating cluster arises, the structure stays static as long as no global parameters are 
changed. There will be a sudden drop in global pressure $P$ as no capillary barriers will have to be overcome
to sustain the global flow rate.   

In the opposite case, clusters of fluid move, fragment and coalesce in an equilibrium process with a global 
pressure fluctuating around a mean value. This is true for the viscous case where 
clusters are not trapped inside a growing percolating cluster. However, we report in this paper 
that this is also the case in a steady state situation even for very low ${\rm Ca}$ which should indicate 
that capillary forces dominate. 

If, on the other hand, ${\rm Ca}$ is large and viscous forces dominate, we will need a much larger
difference in $S$ to get a spanning cluster. This is because trapping of fluid is less likely to
happen and therefore large clusters are more vulnerable to fragmentation through creation of bubbles. 
In the extreme case, where there are no capillary forces at all, a homogeneous mixing will occur.

\subsection{Pressure development}\label{31}

For small ${\rm Ca}$ and subsequently large influence
of capillary forces, the pressure build up is initially slow. A regular structure of the fluid clusters
will decrease the global pressure since the capillary part of the global pressure is dependent of the overall
structure and the single phase pressure $P_s \ll P_c$. 

%%%%%%%%%%%%%%%%%%%%%%%%%%%%%%%%%%%%%%%%%%%%%%%%%%%%%%%%%%%%%%%%%%%%%%%%%%%%%%%%%%%%%%%%%%%%%%%
\begin{figure}[b!]
  \vspace{5mm}
  \includegraphics[width = 0.9\linewidth]{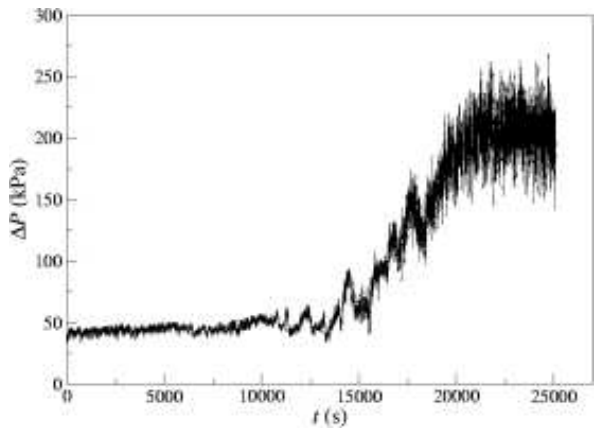}
  \caption{Dynamic evolution of global pressure in a $128\times128$ system at ${\rm Ca} = 3.2 \times 10^{-5}$
    and $M = 1$. In the initial phase of the flow evolution the is a small build up in pressure, but when the
    the wetting front breaks through the initial belt of non wetting fluid the increase in pressure is violent.
    At the last, steady state, regime the fluctuations is approximately five times larger then in the first regime.}
  \label{pressure}
\end{figure}
%%%%%%%%%%%%%%%%%%%%%%%%%%%%%%%%%%%%%%%%%%%%%%%%%%%%%%%%%%%%%%%%%%%%%%%%%%%%%%%%%%%%%%%%%%%%%%%

For low capillary numbers the non-wetting fluid will be blocked in the area around the nodes due to
the capillary barriers of the tubes. Wetting fluid will therefore in large parts of the
system be immobilized. This trapping of fluid gives a lower fractal dimension $\tilde{D}$ of the 
invading IP structure.

For low ${\rm Ca}$, the drainage of the IP part will eventually meet the compact imbibition front of
the wetting fluid due to {\it biperiodic boundary conditions}. 
As long as the source of the drainage transient has an intact ``belt'', as shown in Fig.~\ref{lowCa}, 
in the direction normal to the 
flow direction, the invading front will move through  
bursts in the local pressure. These have a well defined distribution \cite{amhb00}. This regime is seen in Fig.~\ref{pressure}
early in the pressure evolution as a region with a slowly increasing mean pressure. 

When the first non-wetting fingers start to coalesce with the stable displacement of the wetting fluid after
having completed a turn along the torus, a competition between two different processes arises. 
If the percolating non-wetting cluster is not sustained and subsequently
broken, $P_c$ will increase.

Eventually the initial IP cluster will break up and fragment into
smaller clusters as shown in the third picture of Fig.~\ref{lowCa}. 
When there is no initial structure left in the system, the pressure saturates around a mean value with larger
fluctuations than the previous case with the small bursts. 

The pressure development in Fig.~\ref{pressure} can be divided in three different segments. For this case where the
injection rate is slow, the first segment describes an IP regime where the $P_c$ only describes local capillary fluctuations
for the menisci movement along the front. The trapped clusters of wetting fluid will have little effect on the effective
viscosity of the non-wetting fluid and the front is not saturated. 

In the second segment we see a drastic increase in $\Delta P$. Since the system is having the same injection rate and therefore is
governed by capillary effects, this means that the effective dynamic viscosity $\mu_{\rm nw}$ has increased 
and the viscosity ratio $M > 1$.
A qualitative explanation for this increase in viscosity can be seen in connection to the reduced 
permeability of the invaded, wetting
region caused by trapped non-wetting clusters left behind by the compact wetting front.   
The non-wetting front has reached saturation width $w_{\rm s}$.
The result is a more stable front, and it was pointed out by Aker {\it et al} \cite{amh00-2} that $P_c$ is proportional to the height difference
in the front

%%%%%%%%%%%%%%%%%%%%%%%%%%%%%%%%%%%%%%%%%%%%%%%%%%%%%%%%%%%%%%%%%%%%%%%%%%%%%%%%%%%%%%%%
\begin{equation}
  P_c \propto \Delta h^{\kappa} \ ,
  \label{w}
\end{equation}
%%%%%%%%%%%%%%%%%%%%%%%%%%%%%%%%%%%%%%%%%%%%%%%%%%%%%%%%%%%%%%%%%%%%%%%%%%%%%%%%%%%%%%%% 
\noindent where $\kappa = 1.0$. 

In the third and last segment, the flow has reached steady state. The fluctuations in $P_c$ are larger and the mean
value higher due to the unstructured pattern of fluids.

\section{Steady state distributions}\label{4}

In this section we look particularly at the distribution of non-wetting clusters. Because of the capillary barriers
the non-wetting fluid has a much higher tendency to occupy pore-volume near the nodes rather than around the throats 
in the network. 
Due to volume conservation of wetting fluid, this will block the throats and create loop-less fingers of non-wetting
fluid. Moving of trapped clusters must occur, and the steady-state fragmentation and merging
of non-wetting clusters $N(s,t)$ is

%%%%%%%%%%%%%%%%%%%%%%%%%%%%%%%%%%%%%%%%%%%%%%%%%%%%%%%%%%%%%%%%%%%%%%%%%%%%%%%%%%%%%%%%
\begin{equation}
  \partial_t N(s,t) = [\partial_t N]_{\rm merging} + [\partial_t N]_{\rm frag} = 0 \ .
  \label{equil}
\end{equation}
%%%%%%%%%%%%%%%%%%%%%%%%%%%%%%%%%%%%%%%%%%%%%%%%%%%%%%%%%%%%%%%%%%%%%%%%%%%%%%%%%%%%%%%%

\subsection{Distribution of clusters}\label{41}

The distribution of clusters depends on size. 
Fragmentation and merging of clusters in various dimensions are
described i.e. in refs. \cite{gfe95, fbjmos03, rh90}.

At steady-state there will be a typical size $s^*$ of the clusters, and clusters of size
$s \gg s^*$ are very rare.
The distribution of clusters $N(s,L)$ consists of both a regular and a singular part, 
$N(s,L) = N_{\rm reg}(s,L) + N_{\rm sing}(s,L)$, 
and for saturations $S_{\rm nw}$ close to or equal to a critical saturation $S_c$, 
the singular part of $N(s,L)$ will dominate. 

Based on analogy to percolation and finite-size scaling \cite{stauffer94}, 
we assume the general expression for the singular distribution to be 

%%%%%%%%%%%%%%%%%%%%%%%%%%%%%%%%%%%%%%%%%%%%%%%%%%%%%%%%%%%%%%%%%%%%%%%%%%%%%%%%%%%%%%%%
\begin{equation}
  N_{\rm sing}(s,L) \propto L^{\chi} s^{-\tau} f\left( \frac{s}{s^*} \right) \ ,
  \label{ms}
\end{equation}
%%%%%%%%%%%%%%%%%%%%%%%%%%%%%%%%%%%%%%%%%%%%%%%%%%%%%%%%%%%%%%%%%%%%%%%%%%%%%%%%%%%%%%%%

\noindent where $f(x)$ is a cutoff function decaying faster than any power law for $x \gg 1$ and is constant for $x \ll 1$.
The exponent $\chi$ which governs the system size dependence of $N_{\rm sing}$, has the value $\chi = d$ in
pure percolation. We have it here undetermined. 

When the clusters have a fractal dimension $D$, the typical size $s^* \propto l^D$ where 
$l = {\rm min} (L, \xi)$
where $\xi$ is the correlation length. For large system sizes,
the area $A$ the clusters occupy is 

%%%%%%%%%%%%%%%%%%%%%%%%%%%%%%%%%%%%%%%%%%%%%%%%%%%%%%%%%%%%%%%%%%%%%%%%%%%%%%%%%%%%%%%%
\begin{eqnarray}
A & = & \int^{\infty} sN(s,L) {\rm d}s = \int^{\infty} s(N_{\rm reg} + N_{\rm sing}){\rm d}s \nonumber \\
\ & = & A_{\rm reg} + A_{\rm sing} \ ,
\label{A}
\end{eqnarray}
%%%%%%%%%%%%%%%%%%%%%%%%%%%%%%%%%%%%%%%%%%%%%%%%%%%%%%%%%%%%%%%%%%%%%%%%%%%%%%%%%%%%%%%%

\noindent and the singular part scales as

%%%%%%%%%%%%%%%%%%%%%%%%%%%%%%%%%%%%%%%%%%%%%%%%%%%%%%%%%%%%%%%%%%%%%%%%%%%%%%%%%%%%%%%%
\begin{eqnarray}
  A_{\rm sing}  & \sim & L^{\chi}\int^{\infty} s^{1-\tau}f(s/s^*){\rm d}s \nonumber \\
  \            & = & L^{\chi}(s^*)^{2-\tau}\int^{\infty} x^{1-\tau} f(x) {\rm d}x \ .
  \label{Asing}
\end{eqnarray}
%%%%%%%%%%%%%%%%%%%%%%%%%%%%%%%%%%%%%%%%%%%%%%%%%%%%%%%%%%%%%%%%%%%%%%%%%%%%%%%%%%%%%%%%

The lower limit of Eq.~\eqref{Asing} converges when we exclude arbitrary small clusters,
and as a result

%%%%%%%%%%%%%%%%%%%%%%%%%%%%%%%%%%%%%%%%%%%%%%%%%%%%%%%%%%%%%%%%%%%%%%%%%%%%%%%%%%%%%%%%
\begin{equation}
A_{\rm sing} \sim L^{\chi}(s^*)^{2-\tau} = L^{\chi + (2-\tau)D} \ ,
\label{Asing2}
\end{equation}
%%%%%%%%%%%%%%%%%%%%%%%%%%%%%%%%%%%%%%%%%%%%%%%%%%%%%%%%%%%%%%%%%%%%%%%%%%%%%%%%%%%%%%%%
\noindent when we use that $s^* \propto L^D$ near critical saturation.

Long before steady-state, the non-wetting fluid forms IP patterns, and the total area $A \propto s^*$. 
The saturation of the invading, 
non-wetting fluid, then depends on the fractal dimension $D$ and
will scale locally as \cite{lz85}

%%%%%%%%%%%%%%%%%%%%%%%%%%%%%%%%%%%%%%%%%%%%%%%%%%%%%%%%%%%%%%%%%%%%%%%%%%%%%%%%%%%%%%%%
\begin{equation}
S_{nw} \propto L_{loc}^{D-d} \ ,
\label{Snw}
\end{equation}
%%%%%%%%%%%%%%%%%%%%%%%%%%%%%%%%%%%%%%%%%%%%%%%%%%%%%%%%%%%%%%%%%%%%%%%%%%%%%%%%%%%%%%%%

\noindent where $L_{loc}$ is limited to the region where the defending fluid is purely wetting.
Due to the incompressibility of the two phases, trapped
fluids around the hull of the spanning, invading cluster causes the saturation of invading fluid to decrease
even more as the system size $L$ is increased. 

We have that $A = A_{\rm sing}$ locally because this is in the pure IP regime, and there are no fragments of
non-wetting fluid. Following Meakin \cite{m91}, the scaling of the singular cluster surface is

%%%%%%%%%%%%%%%%%%%%%%%%%%%%%%%%%%%%%%%%%%%%%%%%%%%%%%%%%%%%%%%%%%%%%%%%%%%%%%%%%%%%%%%%
\begin{eqnarray}
A_{\rm sing} & = & \int^{\infty} sN_{\rm sing}(s,L) {\rm d}s \nonumber \\ 
\ &\sim & L^d \int^{\infty} s^{1-\tau} f\left( \frac{s}{s^*} \right) {\rm d}s \ ,
\label{Asing3}
\end{eqnarray}
%%%%%%%%%%%%%%%%%%%%%%%%%%%%%%%%%%%%%%%%%%%%%%%%%%%%%%%%%%%%%%%%%%%%%%%%%%%%%%%%%%%%%%%%

\noindent which gives $A_{\rm sing} \sim L^{d+(2 - \tau )D} \sim L^D$ and

%%%%%%%%%%%%%%%%%%%%%%%%%%%%%%%%%%%%%%%%%%%%%%%%%%%%%%%%%%%%%%%%%%%%%%%%%%%%%%%%%%%%%%%%
\begin{equation}
\tau = \frac{D+d}{D} \ .
\label{tau}
\end{equation}
%%%%%%%%%%%%%%%%%%%%%%%%%%%%%%%%%%%%%%%%%%%%%%%%%%%%%%%%%%%%%%%%%%%%%%%%%%%%%%%%%%%%%%%%

In percolation, the fractal dimension of a spanning cluster is $D(p = p_c) = 91/48$ giving 
$\tau \approx 2.05$. 
In the regime where we have invasion percolation the area
of wetting defending clusters behaves as  

%%%%%%%%%%%%%%%%%%%%%%%%%%%%%%%%%%%%%%%%%%%%%%%%%%%%%%%%%%%%%%%%%%%%%%%%%%%%%%%%%%%%%%%%
\begin{eqnarray}
A & = & \int^{\infty} sN(s) {\rm d}s \nonumber \\
\ & \sim & L^D \int^{\infty} s^{1-\tau} f\left( \frac{s}{s^*} \right) {\rm d}s \ .
\label{Awet}
\end{eqnarray}
%%%%%%%%%%%%%%%%%%%%%%%%%%%%%%%%%%%%%%%%%%%%%%%%%%%%%%%%%%%%%%%%%%%%%%%%%%%%%%%%%%%%%%%%

\noindent It was found that for IP structure, $A$ grows as
$L^d$, which imposed a $\tau < 2$ and 
$A \sim L^{D+(2 - \tau )d} \sim L^d$. This gives $\tau = 1 + D/d$ and with a fractal dimension 
$D = 1.82$ with trapped wetting clusters, this gives $\tau = 1.91$ for the distribution
of wetting clusters. 

When steady-state is reached the mean saturation of the two phases stays constant over the entire 
system and every pattern created by the transients are wiped out. We write the non-wetting saturation as

%%%%%%%%%%%%%%%%%%%%%%%%%%%%%%%%%%%%%%%%%%%%%%%%%%%%%%%%%%%%%%%%%%%%%%%%%%%%%%%%%%%%%%%%%%
\begin{equation}
  S_{\rm nw} = \int^{\infty} s n_s {\rm d}s \ ,
  \label{satur}
\end{equation}
%%%%%%%%%%%%%%%%%%%%%%%%%%%%%%%%%%%%%%%%%%%%%%%%%%%%%%%%%%%%%%%%%%%%%%%%%%%%%%%%%%%%%%%%%%
\noindent where $n_s = L^{-d} N(s,L)$. According to Eq.~\eqref{A} we then get contributions to
$S_{\rm nw}$ both from a regular and singular part of $n_s$. Since $S_{\rm nw}$ is constant
and the spanning IP cluster is broken up, the probability that a given 
area-fraction of non-wetting fluid is attached to the spanning cluster changes from unity to

%%%%%%%%%%%%%%%%%%%%%%%%%%%%%%%%%%%%%%%%%%%%%%%%%%%%%%%%%%%%%%%%%%%%%%%%%%%%%%%%%%%%%%%%%%
\begin{equation}
P_{\infty} \propto (S_{\rm nw} - S_c)^{\beta} \sim L^{-\frac{\beta}{\nu}} \quad {\rm for} \quad S_{\rm nw} \geq S_c \ .
\label{Pinf}
\end{equation}
%%%%%%%%%%%%%%%%%%%%%%%%%%%%%%%%%%%%%%%%%%%%%%%%%%%%%%%%%%%%%%%%%%%%%%%%%%%%%%%%%%%%%%%%%%

\noindent We take finite size scaling into account and the correlation length 
$\xi \propto |S - S_c|^{-\nu}$ when $S_{\rm nw}$ is chosen not to be far from $S_c$. 

The singular contribution to $P_{\infty}$ then gives the following relation between critical
exponents

%%%%%%%%%%%%%%%%%%%%%%%%%%%%%%%%%%%%%%%%%%%%%%%%%%%%%%%%%%%%%%%%%%%%%%%%%%%%%%%%%%%%%%%%%%
\begin{equation}
\frac{\beta}{\nu} = (\tau - 2)D + d - \chi \geq 0 \ , 
\label{beta}
\end{equation}
%%%%%%%%%%%%%%%%%%%%%%%%%%%%%%%%%%%%%%%%%%%%%%%%%%%%%%%%%%%%%%%%%%%%%%%%%%%%%%%%%%%%%%%%%%

\noindent since $\beta \geq 0$ and $\nu \geq 0$, which gives 

%%%%%%%%%%%%%%%%%%%%%%%%%%%%%%%%%%%%%%%%%%%%%%%%%%%%%%%%%%%%%%%%%%%%%%%%%%%%%%%%%%%%%%%%%%
\begin{equation}
\tau \geq 2 - \frac{\chi - d}{D} \ . 
\label{tau2}
\end{equation}
%%%%%%%%%%%%%%%%%%%%%%%%%%%%%%%%%%%%%%%%%%%%%%%%%%%%%%%%%%%%%%%%%%%%%%%%%%%%%%%%%%%%%%%%%%
\noindent From this, we can have a value of $\tau < 2$, if the exponent 
$\chi < d$. In percolation where $\chi = 2$ we must have a $\tau > 2$ in order to ensure a 
positive $\beta > 0$.

When we look at the probability $n_s$ of having a cluster of size $s$ in the limit of
$L \rightarrow \infty$ it must approach an asymptotic, constant value. 
Since $n_{s,{\rm sing}} \leq n_s$ and

%%%%%%%%%%%%%%%%%%%%%%%%%%%%%%%%%%%%%%%%%%%%%%%%%%%%%%%%%%%%%%%%%%%%%%%%%%%%%%%%%%%%%%%%%%
\begin{equation}
n_{s,{\rm sing}} \sim L^{\chi - d} s^{-\tau} f\left( \frac{s}{s^*} \right) \rightarrow {\rm Const.} \ L^{\chi - d} \ , 
\label{nsing}
\end{equation}
%%%%%%%%%%%%%%%%%%%%%%%%%%%%%%%%%%%%%%%%%%%%%%%%%%%%%%%%%%%%%%%%%%%%%%%%%%%%%%%%%%%%%%%%%%

\noindent as $L \rightarrow \infty$, and hence $s^* \rightarrow \infty$,
this means that $\chi \leq 2$ for finite $s$. It is clear the our situation differs from
ordinary percolation. 

\subsection{Evolution of wetting clusters}\label{42}
It has been shown both experimentally and in numerical studies that the width of an invading non-wetting front $w_{\rm nw}$
depends both on ${\rm Ca}$ and the time $t$ of the evolution. The front width $w$ of the invading front is calculated
as the standard deviation of number of points $n(y)$ belonging to the front at a distance $y$ from the average position 
of the same front. It has been proposed that $w_{\rm nw}$ scales width time as
 
%%%%%%%%%%%%%%%%%%%%%%%%%%%%%%%%%%%%%%%%%%%%%%%%%%%%%%%%%%%%%%%%%%%%%%%%%%%%%%%%%%%%%%%%
\begin{equation}
w = t^{\beta_d}h(t, {\rm Ca}) \ ,
\label{ws}
\end{equation}
%%%%%%%%%%%%%%%%%%%%%%%%%%%%%%%%%%%%%%%%%%%%%%%%%%%%%%%%%%%%%%%%%%%%%%%%%%%%%%%%%%%%%%%%
where $h(t, {\rm Ca})$ is a crossover function. When the width of the front has reached saturation, i. e.
at large time scales $t \gg w^{1/\beta_d}$, the saturation width $w_{\rm s}$ is no longer dependent of time and
$w_{\rm s}$ is purely a function of ${\rm Ca}$ \cite{fmsh97}. 

\begin{figure}[t!]
\vspace{5mm}
\includegraphics[width = 0.9\linewidth]{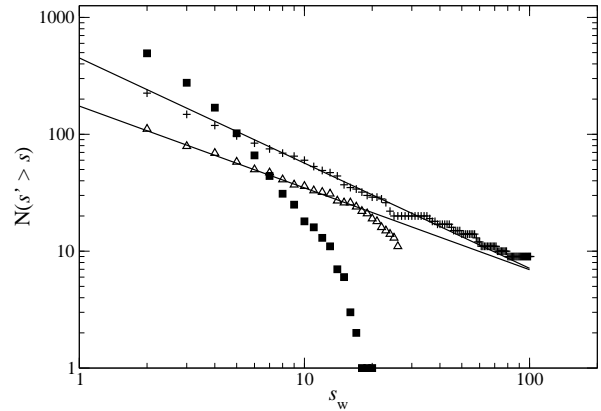}
\caption{Distribution of wetting clusters near the non-wetting front at different stages of evolution. The earliest
stage $(\vartriangle )$ has a slope of $1.71$. The cutoff here is due to lack of larger clusters that will evolve at later stages. As time
increases $(+)$, the slope is now $1.91$, while at steady state $(\blacksquare )$ the power law distribution is completely overrun by the cutoff 
function.}
\label{wetting clusters}
\end{figure}

At low ${\rm Ca}$, the front is wide and trapped clusters of wetting fluid
exists on many different length scales. This situation is best described as
IP with trapping, with a fractal dimension $D = 1.82$ \cite{ffaj86}. The distribution of these wetting clusters
is shown in Fig.~\ref{wetting clusters}. We see a significant difference in the scaling at different times. 
The smallest exponent $\tau$ according to Eq.~\eqref{ms} is
$\tau \approx 1.7$ which is consistent with the experimental findings by Frette {\it et al} \cite{fmsh97}. 
At this point there is a cutoff for large clusters as these have not yet had the time to form. 
In the latter case $\tau \approx 1.9$ which is in more correspondence with the scaling proposed 
by Meakin \cite{m91} in Eq.~\eqref{Awet}. When the system is in steady-state, the wetting clusters are no longer trapped
and the largest wetting clusters will break apart.

%%%%%%%%%%%%%%%%%%%%%%%%%%%%%%%%%%%%%%%%%%%%%%%%%%%%%%%%%%%%%%%%%%%%%%%%%%%%%%%%%%%%%%%%%%%%%%%

%%%%%%%%%%%%%%%%%%%%%%%%%%%%%%%%%%%%%%%%%%%%%%%%%%%%%%%%%%%%%%%%%%%%%%%%%%%%%%%%%%%%%%%%%%%%%%%

\begin{figure*}[t!]
  \begin{minipage}{0.46\linewidth}
    \includegraphics[width = \linewidth]{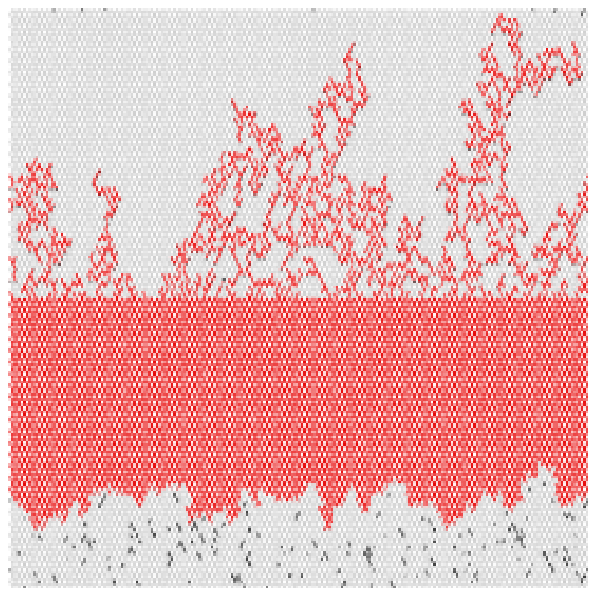}
    \vspace{2mm} 
  \end{minipage}
  \begin{minipage}{0.46\linewidth}
    \includegraphics[width = \linewidth]{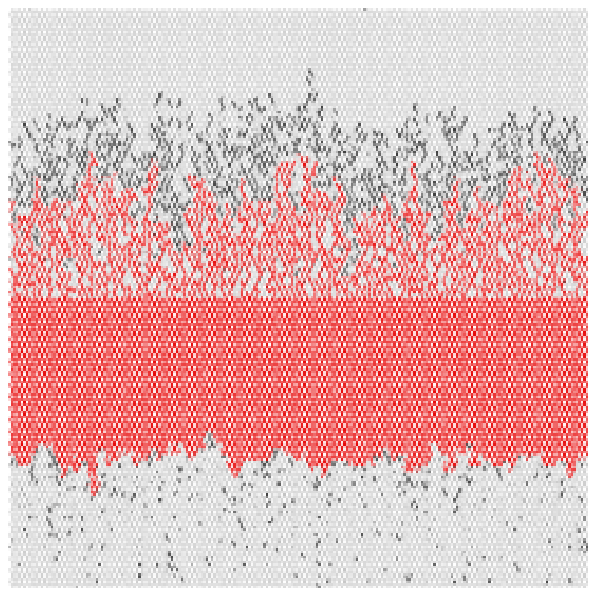}
    \vspace{2mm}
  \end{minipage}
  \begin{minipage}{0.46\linewidth}
    \includegraphics[width = \linewidth]{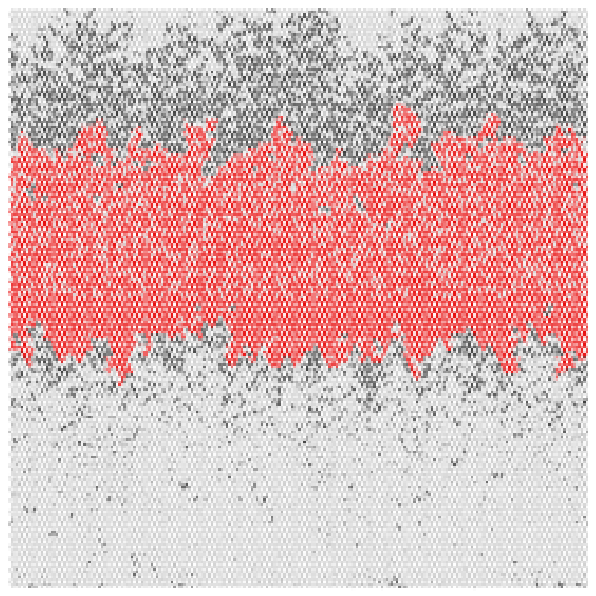}
  \end{minipage}
  \begin{minipage}{0.46\linewidth}
    \includegraphics[width = \linewidth]{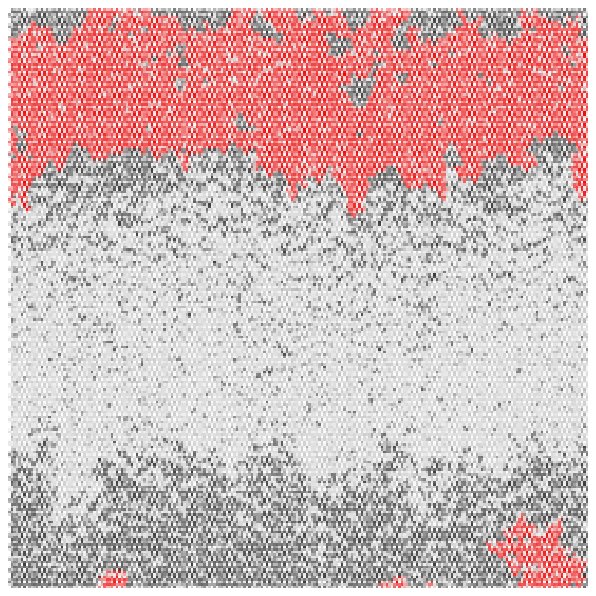}
  \end{minipage}
  \caption{(Color online) Four different snapshots of configurations with different ${\rm Ca}$ taken at approximately the same time. 
    In the upper left figure ${\rm Ca} = 3.2 \times 10^{-5}$. For the next three figures ${\rm Ca}$ is much higher, ${\rm Ca} = 1, 10$ and $100$. The largest connected cluster is colored red in the online version of the article.
    The non-wetting front is now compact with high degree of fragmentation. The viscosity-ratio of the two fluids is $M = 1$, 
    but the effective viscosity may be altered due to decreased permeability caused by fragmentation of the clusters.}
  \label{diffQ}
\end{figure*}

As ${\rm Ca}$ increases, the non-wetting front approaches a qualitatively more compact displacement \cite{w86, fbs05}. 
Looped fingers will also occur since the capillary barrier is weak compared to
and there are no large, immobile wetting clusters as shown in Fig.~\ref{diffQ}.

%%%%%%%%%%%%%%%%%%%%%%%%%%%%%%%%%%%%%%%%%%%%%%%%%%%%%%%%%%%%%%%%%%%%%%%%%%%%%%%%%%%%%%%%%%%%%%%
%%%%%%%%%%%%%%%%%%%%%%%%%%%%%%%%%%%%%%%%%%%%%%%%%%%%%%%%%%%%%%%%%%%%%%%%%%%%%%%%%%%%%%%%%%%%%%%

\subsection{Effect of saturation}\label{43}

For a certain saturation level $S_{\rm nw}$ of non-wetting fluid, at the equilibrium described in Eq.~\eqref{equil}
the non-wetting fluid will percolate the system. We assume as in percolation theory \cite{stauffer94} the 
singular distribution of clusters to be   

%%%%%%%%%%%%%%%%%%%%%%%%%%%%%%%%%%%%%%%%%%%%%%%%%%%%%%%%%%%%%%%%%%%%%%%%%%%%%%%%%%%%%%%%
\begin{equation}
N_{\rm sing}(s) \sim s^{-\tau} \exp \left( -\frac{s}{s^*} \right) \ .
\label{ms2}
\end{equation}
%%%%%%%%%%%%%%%%%%%%%%%%%%%%%%%%%%%%%%%%%%%%%%%%%%%%%%%%%%%%%%%%%%%%%%%%%%%%%%%%%%%%%%%%

Since $s^* \propto \xi^D \propto |S_{\rm nw} - S_c|^{-\nu D}$, we
can write the typical cluster size in means of saturation and get $s^* \propto |S_{\rm nw} - S_c|^{-1/\sigma}$.
The exponent $\sigma = 1/(\nu D)$, and for simplicity we define $c \propto |S - S_c|^{1/\sigma}$. 

We have performed numerical tests of the above presumption with a matching viscosity ratio $M = 1$.
The tests have been initiated with both a complete separation of
wetting and non-wetting fluids and a random mixing of the two phases. The question is whether or not the initial
conditions will affect the steady-state outcome. 

If there exists a fluid configuration that stems from the combined process of drainage and imbibition, the
outcome of the two initial configurations described above should have the same statistical properties. 
However, for low ${\rm Ca}$ we encounter difficulties connected to ``locking'' of immobile configurations due to large 
capillary barriers.
The time-steps in our model become unphysically large. In order to compensate for these effects we apply 
certain shocks to the system where the input flow rate is increased drastically for a small amount of
time steps compared to the total number of ran time steps.

We now consider distributions of clusters obtained from different system sizes and initial configurations.
The clusters are identified through a customized Hoshen-Kopelman algorithm \cite{hk76}. The mass $s$ of the clusters
is a continuous variable, and in order to make a histogram we bin the identifed clusters with a binsize 
$\Delta s = s_{\rm max} \times 1/10000$. The largest cluster $s_{\rm max}$ is five orders of magnitude larger than the unit
size $s = 1$.

As described in previous sections, the case where the two phases are completely separated gives intermediate
configurations with different cluster distributions. The distributions depend on whether the fluid is invading
or withdrawing from a region. In a study by Wagner $et \ al$ \cite{wbmfj97} experiments and simulations of fragmentation
of non-wetting fluid were carried out. The case here was to take a full IP pattern, and reverse the fluid current 
so that the flow regime was changed from drainage to imbibition. A result to note in Ref. \cite{wbmfj97}, is that when
the non-wetting saturation of the IP pattern at breakthrough follows the relation in Eq. \eqref{Snw} as expected, 
the final saturation of non-wetting fluid seemed to be independent of system size. 

In traditional simulation and experiments of imbibition and drainage of different fluids, there is a inlet flow of one phase into
an already fully saturated region of another phase. 
In our simulations, the effective saturation of the two phases are constant throughout the entire experiments. 

The percolation threshold for bond percolation in a regular square lattice is $p_c = 1/2$. 
The distribution of clusters in the system referred to in Fig.~\ref{lowCa} with $S_{\rm{nw}} = 0.5$ has a
clear cutoff for large cluster sizes $s$. It is therefore
clear that -- in order to see clusters of non wetting fluid with size in order of the system size -- we must have a larger
saturation $S_{\rm{nw}}$ and $S_c > 0.5$.

%%%%%%%%%%%%%%%%%%%%%%%%%%%%%%%%%%%%%%%%%%%%%%%%%%%%%%%%%%%%%%%%%%%%%%%%%%%%%%%%%%%%%%%%%%%%%%%
\begin{figure}
  \vspace{5mm}
  \includegraphics[width = 0.9\linewidth]{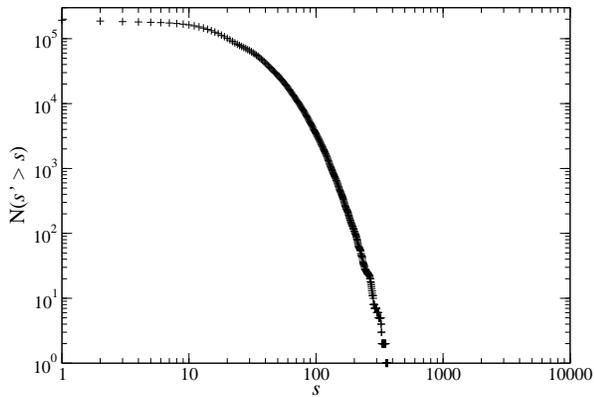}
  \caption{Cumulative distribution $N(s' > s)$ of non-wetting clusters at steady state flow regime 
    for $512 \times 512$ system at ${\rm Ca} = 3.2 \times 10^{-5}$. The results are from a simulation with 
    $S_{nw} = 0.5$ and the clear cutoff at high $s$ indicates that the critical saturation $S_c > 0.5$.}
  \label{512x512clustersCum}
\end{figure}
%%%%%%%%%%%%%%%%%%%%%%%%%%%%%%%%%%%%%%%%%%%%%%%%%%%%%%%%%%%%%%%%%%%%%%%%%%%%%%%%%%%%%%%%%%%%%%% 

\begin{figure}
  \vspace{5mm}
  \begin{minipage}{\linewidth}
    \mbox{
      \includegraphics[width = 0.9\linewidth]{fig7a.eps}}
    \vspace{5mm}
  \end{minipage}
  \begin{minipage}{\linewidth}
    \mbox{
      \includegraphics[width = 0.9\linewidth]{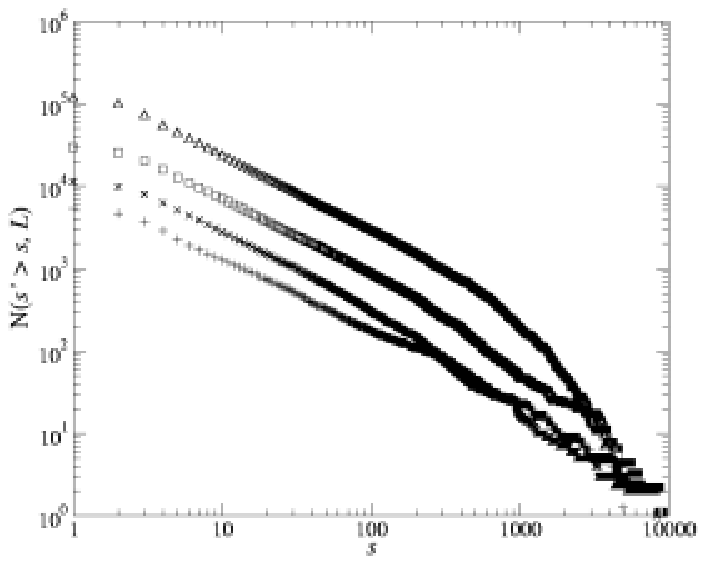}}
  \end{minipage}
  \caption{Log-log plot of distribution of non-wetting clusters at $S_{nw} = 0.69$. The upper figure shows the distribution $N(s)$ 
    for a system of size $1024 \times 1024$. 
    The distribution $N(s,L)$ exhibits a power-law with a slope of $1.93 \pm 0.05$. In the inset, a data-collapse of 
    different system sizes is shown with a size-dependency of $\chi = 1.8$.
    The lower figure shows the cumulative distribution $N(s' > s,L)$ for different system sizes $L\times L$ going from $L = 256$ 
    to $L = 1024$. The distributions have slopes around $0.91 \pm 0.03$. This combined with the data in the upper figure 
    indicates a value of the critical exponent $\tau = 1.92 \pm 0.04$.}
  \label{512x512clusters}
\end{figure}

With a non-wetting saturation $S_{\rm{nw}} = 0.69$ we see a tendency towards a power law distribution of cluster size
$s$. A closer study at the results in Fig. \ref{512x512clusters} reveals a value of the exponent $\tau - 1 = 0.91 \pm 0.03$ from the
cumulative distribution $N(s' > s)$ while a value of $\tau = 1.93 \pm 0.05$ slightly lower than 2 is drawn from the ordinary distribution of
$N(s)$. This is not very far from the prediction of ordinary percolation and in consistence with predictions concerning similar
dynamic models \cite{vf84} which indicate a $\tau$ significantly smaller than 2.

It is hard to pinpoint the exact value of
$S_c$ in a system, but analysis of the cluster moments behaviour 
give us clear indications of the whereabouts of $S_c$. 

%%%%%%%%%%%%%%%%%%%%%%%%%%%%%%%%%%%%%%%%%%%%%%%%%%%%%%%%%%%%%%%%%%%%%%%%%%%%%%%%%%%%%%%%%%%%%%%
%%%%%%%%%%%%%%%%%%%%%%%%%%%%%%%%%%%%%%%%%%%%%%%%%%%%%%%%%%%%%%%%%%%%%%%%%%%%%%%%%%%%%%%%%%%%%%%

\subsection{Scaling analysis}\label{44}
In order to study scaling behavior in the fragmentation and merging process of Eq. \eqref{equil}, we examine
different moments of cluster distribution. 

The $k$th moment is defined as

%%%%%%%%%%%%%%%%%%%%%%%%%%%%%%%%%%%%%%%%%%%%%%%%%%%%%%%%%%%%%%%%%%%%%%%%%%%%%%%%%%%%%%%%
\begin{equation}
M_{k} = \int^{\infty} s^kn_s \ {\rm d}s \ ,
\label{moment}
\end{equation}
%%%%%%%%%%%%%%%%%%%%%%%%%%%%%%%%%%%%%%%%%%%%%%%%%%%%%%%%%%%%%%%%%%%%%%%%%%%%%%%%%%%%%%%%

\noindent and $n_s = n_{s,\rm reg} + n_{s,\rm sing}$, the moments will consist of both a regular and a singular
part. We use percolation methods \cite{stauffer94} to extract information of
critical saturation and critical exponents. 

Since the data we analyze is divided into discrete histogram bins we substitute the integral of Eq.~\eqref{moment} with 
a sum and get for the singular part

%%%%%%%%%%%%%%%%%%%%%%%%%%%%%%%%%%%%%%%%%%%%%%%%%%%%%%%%%%%%%%%%%%%%%%%%%%%%%%%%%%%%%%%%
\begin{equation}
M_{k, {\rm sing}} = \sum_s s^k n_s \propto L^{\chi - 2} c^{\tau - 1 - k} \ ,
\label{momentSing}
\end{equation}
%%%%%%%%%%%%%%%%%%%%%%%%%%%%%%%%%%%%%%%%%%%%%%%%%%%%%%%%%%%%%%%%%%%%%%%%%%%%%%%%%%%%%%%%
\noindent with the assumption that $\chi - 2 < 0$, where $\chi$ was defined in Eq.~\eqref{ms}.

The total $k$th moment is \cite{gfe95}

%%%%%%%%%%%%%%%%%%%%%%%%%%%%%%%%%%%%%%%%%%%%%%%%%%%%%%%%%%%%%%%%%%%%%%%%%%%%%%%%%%%%%%%%
\begin{equation}
M_{k} = A_k + B_k c + C_k c^{\tau - 1 - k} \ ,
\label{momentTot}
\end{equation}
%%%%%%%%%%%%%%%%%%%%%%%%%%%%%%%%%%%%%%%%%%%%%%%%%%%%%%%%%%%%%%%%%%%%%%%%%%%%%%%%%%%%%%%%

\noindent where the last term stems from the singular part and dominates for $k \geq 2$, since
the finite-size contribution from Eq. \eqref{momentSing} is small. 
 
In particular, the second moment $M_2$ is dominated by $M_{2, {\rm sing}}$  and

%%%%%%%%%%%%%%%%%%%%%%%%%%%%%%%%%%%%%%%%%%%%%%%%%%%%%%%%%%%%%%%%%%%%%%%%%%%%%%%%%%%%%%%%
\begin{eqnarray}
    M_2 &\propto& M_{2,\rm sing} = C_2 L^{\chi - 2}c^{\tau - 3} \nonumber \\
    &\propto &|S - S_c|^{-\frac{3-\tau}{\sigma} - (\chi - 2) \nu}  \nonumber \\
    \   & = &|S - S_c|^{-\tilde{\gamma}} \ ,
  \label{2ndmoment}
\end{eqnarray}
%%%%%%%%%%%%%%%%%%%%%%%%%%%%%%%%%%%%%%%%%%%%%%%%%%%%%%%%%%%%%%%%%%%%%%%%%%%%%%%%%%%%%%%%

\noindent when $s^* \sim L^D$ near critical saturation. Then $(\chi - 2) \nu$ gives a correction to the exponent 
$\gamma = ( 3-\tau )/\sigma$ from ordinary percolation theory, and 
$\tilde{\gamma} = \gamma + (\chi - 2) \nu$.

\begin{figure}
\vspace{5mm}
\includegraphics[width = 0.9\linewidth]{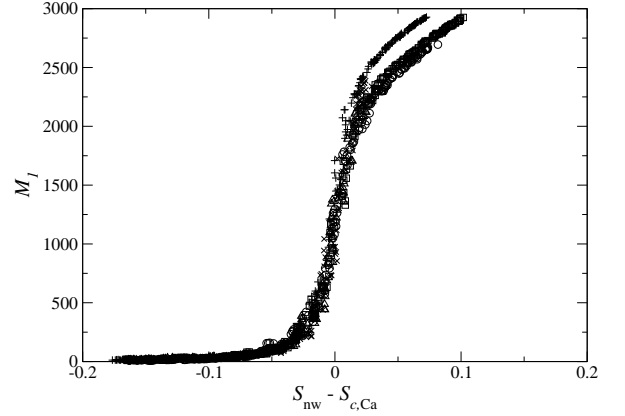}
\caption{First moment of the largest cluster plotted as a function of saturations $S_{nw}$ for $L = 512$. The three curves are for hence
${\rm Ca} = 3.2 \times 10^{-3}$, ${\rm Ca} = 3.2 \times 10^{-4}$ and $ca = 3.2 \times 10^{-5}$. The percolation thresholds are shifted towards higher
values of $S_{nw}$ as ${\rm Ca}$ increases. The transitions are located at approximately $S_c = 0.675$, $S_c = 0.70$ and $S_c = 0.725$ and hence 
collapsed.}
\label{2moments}
\end{figure}

Results from different calculations of $M_2$ for different values of ${\rm Ca}$ are shown in Fig.~\ref{2moments}.
We notice that there are two regimes in the plot separated by a crossover at a certain value of $|S_{\rm nw} - S_c|$.

This can be understood finite-size scaling, where we near a critical density $p_c$
apply the finite-size ansatz 

%%%%%%%%%%%%%%%%%%%%%%%%%%%%%%%%%%%%%%%%%%%%%%%%%%%%%%%%%%%%%%%%%%%%%%%%%%%%%%%%%%%%%%%%
\begin{equation}
\Pi = \Phi [(p - p_c)L^{1/\nu}] \ .
\label{ansatz1}
\end{equation}
%%%%%%%%%%%%%%%%%%%%%%%%%%%%%%%%%%%%%%%%%%%%%%%%%%%%%%%%%%%%%%%%%%%%%%%%%%%%%%%%%%%%%%%%
\noindent This implies that the effective, critical density $p_{\rm eff}$ is

%%%%%%%%%%%%%%%%%%%%%%%%%%%%%%%%%%%%%%%%%%%%%%%%%%%%%%%%%%%%%%%%%%%%%%%%%%%%%%%%%%%%%%%%
\begin{equation}
p_{\rm eff} = \int p \left( \frac{{\rm d}\Pi}{{\rm d}p} \right) \ {\rm d}p \ .
\label{ansatz2}
\end{equation}
%%%%%%%%%%%%%%%%%%%%%%%%%%%%%%%%%%%%%%%%%%%%%%%%%%%%%%%%%%%%%%%%%%%%%%%%%%%%%%%%%%%%%%%%

Hence, in our case we use this expression with our saturation for the variable $p$ and get

%%%%%%%%%%%%%%%%%%%%%%%%%%%%%%%%%%%%%%%%%%%%%%%%%%%%%%%%%%%%%%%%%%%%%%%%%%%%%%%%%%%%%%%%
\begin{equation}
S_{{\rm eff}} = S_c + \frac{A}{L^{1/\nu}} \ . 
\label{ansatz3}
\end{equation}
%%%%%%%%%%%%%%%%%%%%%%%%%%%%%%%%%%%%%%%%%%%%%%%%%%%%%%%%%%%%%%%%%%%%%%%%%%%%%%%%%%%%%%%%
\noindent Setting this new critical saturation in for $S_c$ in Eq. \eqref{2ndmoment} we see the
two regimes as

%%%%%%%%%%%%%%%%%%%%%%%%%%%%%%%%%%%%%%%%%%%%%%%%%%%%%%%%%%%%%%%%%%%%%%%%%%%%%%%%%%%%%%%%
\begin{equation}
  M_2 \propto 
  \begin{cases}
    |S_{\rm nw} - S_c|^{\tilde{\gamma}} & {\rm for} \ |S_{\rm nw} - S_c| \gg \frac{A}{L^{1/\nu}},\\
    {\rm Const.} & {\rm for} \ |S_{\rm nw} - S_c| \ll \frac{A}{L^{1/\nu}},
  \end{cases} 
  \label{ansatz4}
\end{equation}
%%%%%%%%%%%%%%%%%%%%%%%%%%%%%%%%%%%%%%%%%%%%%%%%%%%%%%%%%%%%%%%%%%%%%%%%%%%%%%%%%%%%%%%%
\noindent when only $S_{\rm nw}$ varies and $L$ is constant.

%%%%%%%%%%%%%%%%%%%%%%%%%%%%%%%%%%%%%%%%%%%%%%%%%%%%%%%%%%%%%%%%%%%%%%%%%%%%%%%%%%%%%%%%%%%%%%%
%\onecolumngrid

%%%%%%%%%%%%%%%%%%%%%%%%%%%%%%%%%%%%%%%%%%%%%%%%%%%%%%%%%%%%%%%%%%%%%%%%%%%%%%%%%%%%%%%%%%%%%%%

\begin{figure}[b]
\vspace{5mm}
\includegraphics[width = 0.9\linewidth]{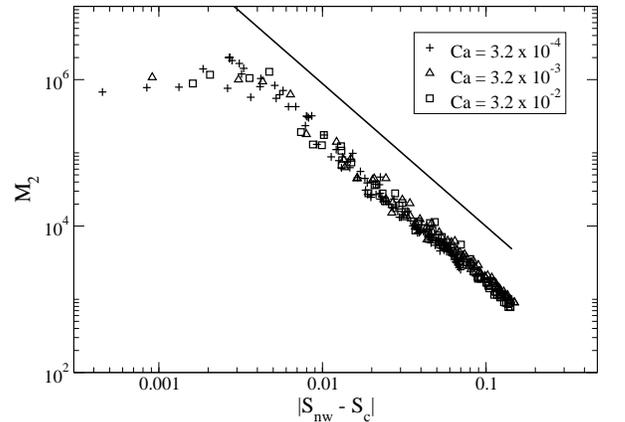}
\caption{Log-log plot of the second moment for different ${\rm Ca}$ for $L = 512$.
The data are collapsed using the values for $S_c$ obtained previously. There is only a narrow 
range where the power-law is valid due to finite-size effects. For $S_{\rm nw} < S_c$ far away from
$S_c$ we see a cut-off and for $|S_{\rm nw} - S_c| \ll 1/L^{1/\nu}$ the function approaches a constant.
However, the obtained value $\tilde{\gamma} = 1.95 \pm 0.06$ is reasonable. 
}
\label{2momentsLog}
\end{figure}

The critical saturations are moving towards larger values as ${\rm Ca}$ increases. Since the fluctuations around
the critical saturations are large, it is difficult to establish accurate value of $S_c$. 
However, Fig.~\ref{2momentsLog} with the different values of $S_{\rm eff}$ obtained in Fig.~\ref{2moments},
gives a power law behaviour with $\tilde{\gamma} = 1.95 \pm 0.06$. This value is significantly lower than
the percolation value of $\gamma$ but this can be understood partly from the system size correction
${\chi - 2} \neq 0$.

%%%%%%%%%%%%%%%%%%%%%%%%%%%%%%%%%%%%%%%%%%%%%%%%%%%%%%%%%%%%%%%%%%%%%%%%%%%%%%%%%%%%%%%%%%%%%%%

%%%%%%%%%%%%%%%%%%%%%%%%%%%%%%%%%%%%%%%%%%%%%%%%%%%%%%%%%%%%%%%%%%%%%%%%%%%%%%%%%%%%%%%%%%%%%%%

Since the critical value of saturation of non-wetting fluid decreases with the capillary number, we expect more
or less an approach towards percolation for extremely low ${\rm Ca}$. On the other hand, when viscous forces are totally
dominant and ${\rm Ca} \rightarrow \infty$, the expected threshold should be around unity. This expectation
is based on the presumption that tiny bubbles of either phase will be overrepresented and effectively break up large clusters forming
out of one fluid phase. Then no spanning non-wetting cluster can form even at very high $S_{\rm nw}$.

\begin{figure}
\includegraphics[width = 0.9\linewidth]{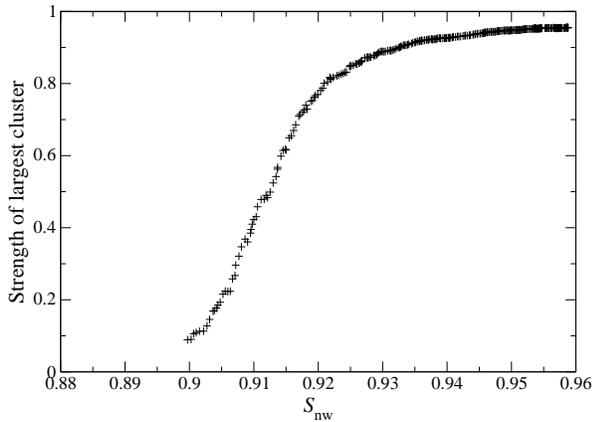}
\caption{Strength of largest non-wetting cluster for ${\rm Ca} \rightarrow \infty$. The data indicates a threshold at
$S_{\rm nw} = 0.91$ which is lower than unity. We can explain this by a lower cutoff of small clusters which will be of
significance when operating with no surface tension. The data are from systemsize $L = 512$ and taken from the mean of ten
different runs.}
\label{infty}
\end{figure}

It is of course a matter of view how small we accept such bubbles to be, because at
some point we must regard nodes to be connected even though microscopic fragments of wetting fluid separates them. 
These microscopic clusters, which we disregard in our calculations, do have an effect on $S_c$ as can be seen in Fig~\ref{infty}.
Here we see $S_c = 0.91$ which is significantly lower than unity.

%%%%%%%%%%%%%%%%%%%%%%%%%%%%%%%%%%%%%%%%%%%%%%%%%%%%%%%%%%%%%%%%%%%%%%%%%%%%%%%%%%%%%%%%%%%%%%%
%%%%%%%%%%%%%%%%%%%%%%%%%%%%%%%%%%%%%%%%%%%%%%%%%%%%%%%%%%%%%%%%%%%%%%%%%%%%%%%%%%%%%%%%%%%%%%%

In order to find a value for $\sigma$, we use the ratio 

%%%%%%%%%%%%%%%%%%%%%%%%%%%%%%%%%%%%%%%%%%%%%%%%%%%%%%%%%%%%%%%%%%%%%%%%%%%%%%%%%%%%%%%%
\begin{equation}
\frac{M_{k+1}}{M_{k}} = \frac{1}{c}, \quad {\rm for} \quad k \geq 2 \ . 
\label{momentRatio}
\end{equation}
%%%%%%%%%%%%%%%%%%%%%%%%%%%%%%%%%%%%%%%%%%%%%%%%%%%%%%%%%%%%%%%%%%%%%%%%%%%%%%%%%%%%%%%%

\noindent Then we can easily find
an estimate for $\sigma$, and even though there is a finite-size correction to the scaling of $M_k$, this
will vanish when we measure the ratio in Eq. \eqref{momentRatio}.

\begin{figure}
\includegraphics[width = 0.9\linewidth]{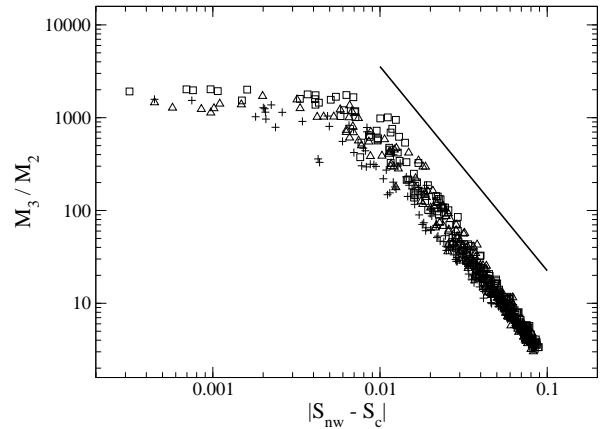}
\caption{Log-log plot of the ratio $M_3 / M_2$ for $L = 512$. The solid line has a slope of $-2.14$ and
gives $1/\sigma = 2.14\pm 0.08$ using Eq.~\eqref{momentRatio}. This gives $\sigma = 0.47\pm0.08$.}
\label{sigmafig}
\end{figure}

The measured value of $\sigma$ from the relation $c \propto |S_{\rm nw} - S_c|^{1/\sigma}$ is shown in Fig.~\ref{sigmafig}.
We find $\sigma = 0.47 \pm 0.08$, but this value is dependent of the chosen
value of the critical saturation $S_c$. This value is arguable. However, the ratio of $\tilde{\gamma}$ and $\sigma$ is assumed to
be constant, and hence the value of $\tau$ in our simulations. Since $D = 1/\sigma \nu$ we get the relation

%%%%%%%%%%%%%%%%%%%%%%%%%%%%%%%%%%%%%%%%%%%%%%%%%%%%%%%%%%%%%%%%%%%%%%%%%%%%%%%%%%%%%%%%
\begin{equation}
\tilde{\gamma} \sigma = 3 - \tau + \frac{\chi - 2}{D} \ .
\label{gammaSigma}
\end{equation}
%%%%%%%%%%%%%%%%%%%%%%%%%%%%%%%%%%%%%%%%%%%%%%%%%%%%%%%%%%%%%%%%%%%%%%%%%%%%%%%%%%%%%%%%

There is of course uncertanties connected to the size effects related to $\chi - 2$, but with the
data-collapse in Fig.~\ref{512x512clusters} which suggested a $\chi - 2 = -0.2$ and a fractal dimension for
the non-wetting clusters in the area $D = 1.82 - 1.89$ \cite{fmsh97, wbmfj97}, the relation above should
give a $\tau \approx 1.97$. This is consistent with our measured $\tau = 1.92 \pm 0.04$.

%%%%%%%%%%%%%%%%%%%%%%%%%%%%%%%%%%%%%%%%%%%%%%%%%%%%%%%%%%%%%%%%%%%%%%%%%%%%%%%%%%%%%%%%%%%%%%%

%%%%%%%%%%%%%%%%%%%%%%%%%%%%%%%%%%%%%%%%%%%%%%%%%%%%%%%%%%%%%%%%%%%%%%%%%%%%%%%%%%%%%%%%%%%%%%%

\section{Conclusion}\label{5}

In this paper we have presented results of cluster distribution in a two-phase flow network simulator
in two dimensions. Different to most studies of porous flow simulations in networks, we apply 
{\it biperiodic boundary conditions}. This allows us to study steady-state conditions
which are more similar to those encountered deep inside natural reservoirs. 

The simulations show that there is a crossover from unstable front propagation to a compact steady-state 
flow as the patterns formed by initial transients are broken up. From the pressure development in Fig.~\ref{pressure} 
this change in flow regime will result in an increase in $\Delta P$ and a more violent evolution
of burst avalanches. We believe that the fragments of non-wetting fluid sustain their fractal properties, but
the $S_{nw}$ at which we get a spanning non-wetting clusters is dependent of ${\rm Ca}$ even though the fractal front is
completely dissolved. 

The steady-state distribution of clusters is shown to share many characteristics with that of ordinary percolation
when it comes to critical properties and the extraction of critical exponents. 
We find a power-law distribution of non-wetting clusters near critical saturation. 
This is obtained regardless of the initial
organization of fluids outside the steady-state regime. Even though $\tau = 1.92 \pm 0.04$ is slightly lower than the value for pure percolation
it is consistent with results obtained from other dynamical models and fragmentation studies \cite{kh96}. 
From the order parameter $P_{\infty} \propto (S - S_c)^{\beta}$, non-negativity of the exponent $\beta > 0$ may be ensured even 
though $\tau < 2$. This is shown through a finite-size analysis.

\begin{acknowledgements}
We thank Martin Ferer, Henning Knudsen and Knut Jørgen Måløy for very helpful discussions. 
This work has been financed through Norwegian Research Council (NFR) Grant No. 154535/432 
and parts of the computations were done on the Norwegian HPC Program NOTUR.
\end{acknowledgements}

\end{document}